# Magnetic-field-induced rotation of light with orbital angular momentum

Shuai Shi,[1,2] Dong-Sheng Ding,[1,2,a)] Zhi-Yuan Zhou,[1,2] Yan Li,[1,2] Wei Zhang,[1,2] and Bao-Sen Shi[1,2,b)]

[1]*Key Laboratory of Quantum Information, University of Science and Technology of China, Hefei, Anhui 230026, China*

[2]*Synergetic Innovation Center of Quantum Information & Quantum Physics, University of Science and Technology of China, Hefei, Anhui 230026, China*

Light carrying orbital angular momentum (OAM) has attractive applications in the fields of precise optical measurements and high capacity optical communications. We study the rotation of a light beam propagating in warm $^{87}$Rb atomic vapor using a method based on magnetic-field-induced circular birefringence. The dependence of the rotation angle on the magnetic field makes it appropriate for weak magnetic field measurements. We quote a detailed theoretical description that agrees well with the experimental observations. The experiment shown here provides a method to measure the magnetic field intensity precisely and expands the application of OAM-carrying light. This technique has advantage in measurement of magnetic field weaker than 0.5 Gauss, and the precision we achieved is 0.8 mGauss.

___________________________

a) Electronic mail: dds@ustc.edu.cn

b) Electronic mail: drshi@ustc.edu.cn

The light-carrying spatial modes with a helical phase front of $\exp(il\alpha)$ have become interesting and important since Allen et al. recognized in 1992 that these modes carried a well-defined orbital angular momentum (OAM).[1] With further developments,[2,3] light with OAM can in principle carry an arbitrarily large amount of information, increasing the information capacity of optical communication networks for both classical and quantum optical communications,[4,5] due to the inherent infinite degrees of freedom for OAM. The interaction between helical light beams and matter have yielded tools such as optical tweezers and optical spanners.[6,7] Recently, methods have been developed to measure various physical parameters involving OAM-carrying light. For example, Lavery et al. have detected the angular frequency of a spinning object by analyzing the angular Doppler frequency shifts of incident light beams with opposite OAM.[8] Our previous work provides a method for measuring the temperature of birefringent crystals with high precision.[9] The "fan-like" interference pattern is used to measure the polarization and OAM of the pump laser in a four-wave mixing process.[10,11]

Here, we develop a method to measure the magnetic field intensity using a combination of OAM-carrying light beams and circular birefringence of warm $^{87}$Rb atomic vapor subjected to an applied magnetic field. We can determine the magnetic field intensity by analyzing the "fan-like" pattern of the rotation angle. Varying the magnetic field intensity results in an observable rotation of the pattern. From the dispersive nature of the medium, the pattern obtained under uniform changes in intensity rotates rapidly when the field is weak and slowly when the field is strong, making it suitable for both strong and weak magnetic field measurements. Moreover, the rotational direction reverses twice as the magnetic field polarity changes, so that we can determine the zero-point of the-magnetic field from the two-reversal points of the rotation-direction.

The experimental setup is shown in Fig. 1. We describe the transformation of the light beam using the language of quantum mechanics.

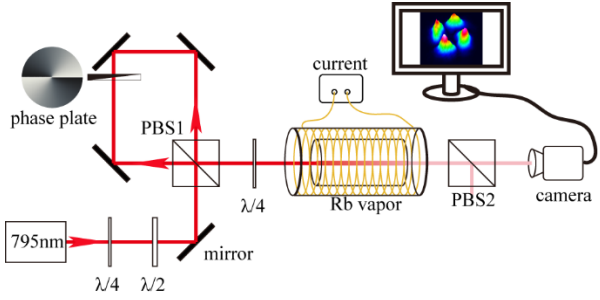

FIG. 1. Experimental setup. The phase plate shows optical thickness increases with azimuthal position. The yellow helix surrounding the Rb vapor represents the electromagnetic coil. The screen shows the intensity distribution of the interference pattern.

A Sagnac configuration is utilized to generate an initial hybrid-superposition of OAM and polarization. The 795-nm laser beam from an external cavity diode laser (Toptica DL100) passes through a quarter-wave plate (QWP) and a half-wave plate and becomes 45° linearly polarized. Since the polarization beam splitter (PBS1) allows horizontal (H) and vertical (V) polarized components pass through the spiral phase plate oppositely. And the thickness of the spiral phase plate increases clockwise and anticlockwise along azimuthal angle for opposite directions. The H and V components obtain an OAM of $l = -2, 2$, respectively. Subsequently, the QWP converts the H and V polarizations to left (L) and right (R) circular polarizations, respectively. We have then prepared the initial state [12-14]:

$$|\Phi\rangle_0 = \frac{1}{\sqrt{2}}\left(|L\rangle \otimes |l\rangle + |R\rangle \otimes |-l\rangle\right). \tag{1}$$

Here $|L\rangle$ and $|R\rangle$ represent the left and right circular polarization states of the light, respectively; $|l\rangle$ represents the OAM quantum state. Equation 1 describes a vector beam,[15-17] which in the language of quantum mechanics is a hybrid-superposition, a superposition state of different degrees of freedom (in this instance spin and OAM). Afterwards, this helically light beam propagates through warm $^{87}$Rb atomic vapor, the temperature of which is stabilized at 89°C, and the atomic density of cell is about $7.23\times10^{12} cm^{-3}$. The signal become more sensitive to magnetic field intensity, but the absorption also enhanced, as the temperature increase. The power of the applied light is 21 mW, and the laser beam diameter is 3 mm. The longitudinal and transverse background magnetic field intensities are $0.20\pm0.02$ and $0.14\pm0.03$ Gauss, respectively. A home-made electromagnetic coil is used to apply a longitudinal magnetic field to the warm atomic vapor. We use a resistance wire as a slide rheostat to adjust the coil current to obtain precise settings of the weak magnetic field intensity. The ratio of the magnetic field intensity to the current value is determined by a DC magnetometer (Alphalab). The 795-nm laser frequency is tuned to the $5S_{1/2}$ F=2 → $5P_{1/2}$ F=1 transition of $^{87}$Rb, which is determined by saturated-absorption spectroscopy. There are two reasons for the choice of this transition; first, as the applied magnetic field intensity increases, the sublevels of $5P_{1/2}$ will never cross, and second, the correspondent sublevels are most similar to the theoretical model.

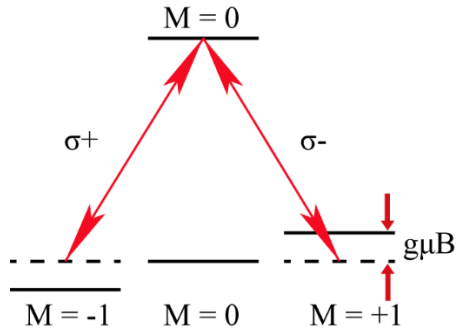

FIG. 2. Transition diagram. The Zeeman sublevels of the ground states are shifted when a longitudinal magnetic field is applied.

When a longitudinal magnetic field is applied, the Zeeman sublevels are shifted (Fig. 2) and the resonance frequencies for the two circular polarizations are different. As a result, they experience different refractive indices, $n_L$ and $n_R$, respectively.[18] This difference in refractive indices results in a relative phase shift $\Delta\varphi = (2\pi/\lambda)(n_L - n_R)d$ between the two components, where d is the propagation distance in the atomic vapor. Hence, after the light beam passes through the $^{87}$Rb vapor cell, the state of the light changes from $|\Phi\rangle_0$ to

$$|\Phi\rangle_1 = \frac{1}{\sqrt{2}}\left(e^{i\Delta\varphi/2}|L\rangle\otimes|l\rangle + e^{-i\Delta\varphi/2}|R\rangle\otimes|-l\rangle\right) \quad (2)$$

PBS2 allows the H component of the L and R circular polarizations to pass through and subsequently interfere. The final state is written

$$|\Phi\rangle_2 = \frac{1}{2}\left(e^{i\Delta\varphi/2}|l\rangle + e^{-i\Delta\varphi/2}|-l\rangle\right)\otimes|H\rangle \quad (3)$$

Since the phase of the helical light beam only depends on the azimuthal angle $\alpha$, and its amplitude only depends on the distance to the optic axis $r$.[1] Hence, the state corresponding to helical light can be written as $\exp(il\alpha)E(r)$, where E(r) is the amplitude distribution of the OAM state and therefore, the state $|\Phi\rangle_2$ can be written:

$$|\Phi\rangle_2 = \cos l(\theta + \alpha) E(r)|H\rangle \quad (4)$$

where $\theta = \Delta\varphi/2l$. Equation (4) shows the dependence of the amplitude on azimuthal angle $\alpha$ for the state. As $\alpha$ ranges from 0 to $2\pi$, there are $2|l|$ zero amplitude angular positions corresponding to dark lines; $\theta$ is the angle associated with the rotation in the interference pattern when the magnetic field is applied; the pattern rotates as the intensity changes. Because the absorption in warm atomic vapor changes as the applied magnetic field is varied, measuring the Faraday rotation of linear polarization requires rotating the polarization analyzer or measuring different polarizations of the beam.[19] Compared with this method, our method has the advantage that we can monitor the rotation in real-time, record several interference patterns in very short time, and improve the accuracy by using a high-resolution camera.

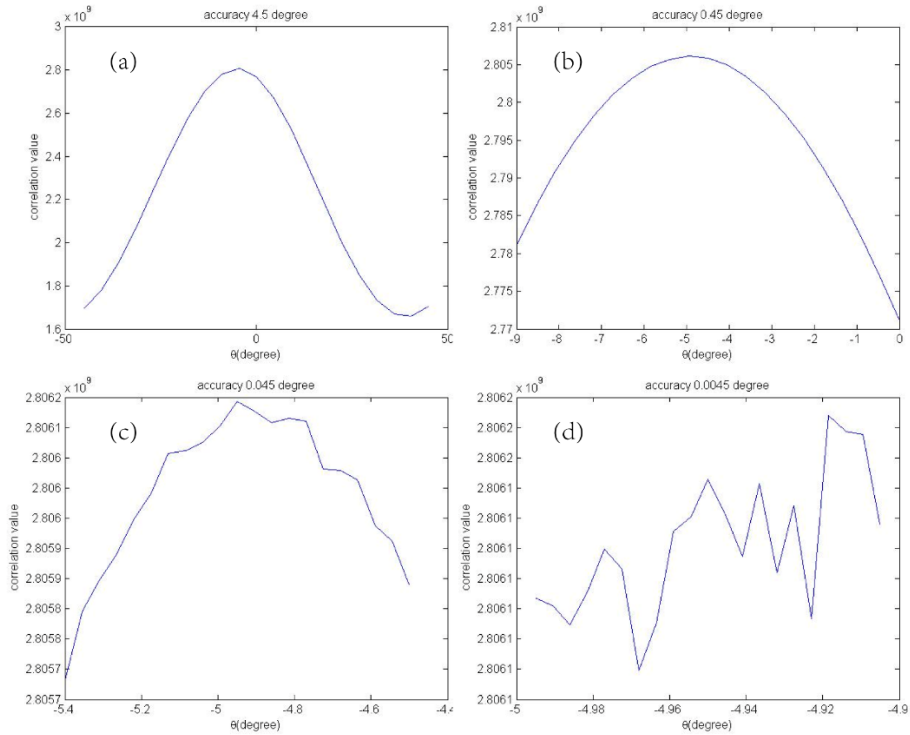

Fig.3. The typical correlation curves of different accuracy. (a) accuracy of 4.5 degree. (b) accuracy of 0.45 degree. (c) accuracy of 0.045 degree. (d) accuracy of 0.0045 degree.

We obtain the rotation angle by calculating the correlation value of the two interference patterns. By scanning the rotation angle of the reference image in small steps and calculating the correlation value with respect to the target image, we can obtain the correlation curve as a function of rotation angle (Fig.3). If the correlation value reaches a maximum, the value along the abscissa is the angle of rotation between the reference and target images. From the figures, the reliable accuracy of this measurement method is nearly 0.045 degrees, and we can further improve the accuracy by increasing the resolution of the interference pattern.

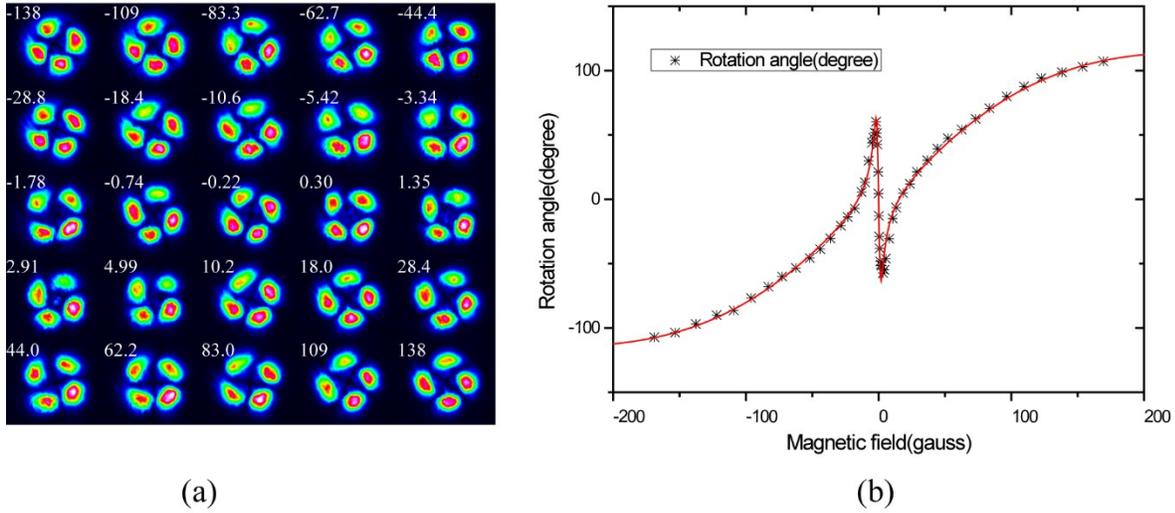

(a)                                         (b)

FIG. 4. (a) Interference patterns for different magnetic field; the sign and number signify direction and intensity (Gauss) of the magnetic field. (b) Variation of rotation angle with magnetic field; the solid line is the fitted curve given by Eq. (5), and the asterisks represent the experimental data.

Figure 4(a) depicts the interference patterns for $l=2$ that we obtained from the experiments with different magnetic field ranging from −138 to +138, the sign and number signify direction and intensity (Gauss) of the magnetic field. As the intensity changes uniformly, the interference pattern rotates slowly if the magnetic field is strong and rotates rapidly if the magnetic field is weak. Moreover, the rotational direction reverses twice as the polarity of the magnetic field alternates. We find a characteristic dispersion-like shape on measuring the dependence of the rotation angle on the magnetic field intensity (Fig. 4(b)). The slope of the curve determines the sensitivity of the rotation angle on the magnetic field intensity. The slope of the curve is about 59°/Gauss for weak magnetic fields and hence points to a means of measuring such fields, taking into account the accuracy of the angular measurement, the corresponding precision of the measurement of weak magnetic field is 0.8 mGauss. There are two principal mechanisms for the steep variation in refractive index. One is the velocity-selective modifications of the atomic population distributions by narrow-band light; the other is related to light-induced coherences between Zeeman sublevels.[18] The center of symmetry of the curve corresponds to the zero-point of the-magnetic-field, assuming the atomic gas is isotropic. There is a relative angle (12.7204 degree) between the interference patterns of the zero-point of the-magnetic-field and the zero-point of the coil current. And the slope of the curve is about 59°/Gauss for weak magnetic fields. The background magnetic field intensity is calculated to be $0.2156 \pm 0.0008$ Gauss, which agrees well with the strength of Earth's magnetic field. The theory of nonlinear magneto-optical rotation near resonance for alkali atoms can be used to explain this phenomenon.[18, 20-22] There are three assumptions for the theoretical model: The atoms that leave the laser beam have their polarization destroyed before entering the beam again; The light-intensity profile is uniform, and model the atom's transit through the beam by assuming a uniform relaxation rate $\gamma$ equal to the inverse of the average transit time. As the complexity of a system increases, the complexity of the analytical solutions grows extremely rapidly.

The expression for $\theta$ is

$$\theta = \frac{6\Gamma\Omega_L d}{lDd_0}\{8\Omega_L^2[8\Omega_L^2 + \Gamma^2(\kappa_2 + 2) - 8\Delta^2] \\ + \gamma[4\gamma(\Gamma^2 - 4\Delta^2) - \Gamma^3\kappa_2(\kappa_2 + 2)]\} \tag{5}$$

The denominator D is:

$$D = 8\Omega_L^2[32\Gamma^2(\kappa_2 + 3)(\Omega_L^2 + \Delta^2) + 192(\Delta^2 - \Omega_L^2)^2 \\ + \Gamma^4(\kappa_2 + 2)(\kappa_2 + 6)] + 2\gamma^2[\Gamma^2(\kappa_2 + 2)^2 + 16\Delta^2] \\ [\Gamma^2(\kappa_2 + 3) + 12\Delta^2]$$

where $\Omega_L = g\mu_0 B/\hbar$ ($\Omega_L = 0.7\mathrm{MHz}/\mathrm{G}$) is the Larmor frequency, Since the Larmor frequency depends on the magnetic field intensity ($B$) and the Landé g-factor ($g$) for the lower atomic state. So more precision might be obtained on a transition with greater Landé g-factor. $\Delta = \omega - \omega_0$ ($\Delta=0\text{MHz}$) the detuning from resonance, $\kappa_2=\Omega_R^2/\Gamma\gamma$ ($\kappa_2=3.3$) the optical-pumping saturation parameter, $l$ the OAM of the light beam, $\Omega_R$ the optical Rabi frequency, $\Gamma$ ($\Gamma=266\text{MHz}$) the spectral width of the absorption line, $d_0$ ($d_0=0.5\text{cm}$) the unsaturated absorption length on resonance, $d$ ($d=5\text{cm}$) the length of the Rb-vapor cell, and $\gamma$ ($\gamma=0.004\Gamma$) the rate at which atoms fly through the light beam. The nonlinear effect stems from the evolution of atomic polarization. The optically pumped atoms evolve in a longitudinal magnetic field, and relax after an average time $1/\gamma$. If the magnetic field is high enough that atoms can process a full revolution before relaxing, the atomic polarization begins to average out, and as a result, the rotation direction of the pattern reverses. Equation (5) explains the dispersion-like shape of the curve very well.

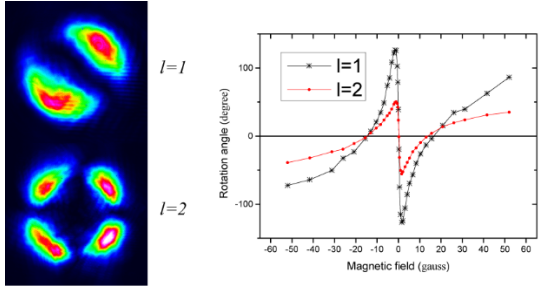

Fig.5. Variation of rotation angle with magnetic field for light carrying OAM $l$=1,2.

Finally, we compare the rotation angle of light carrying OAM l=1 with that of light carrying OAM l=2. It is consistent with our explanation that the rotation angle of l=1 is twice as that of l=2.

In summary, we used a Sagnac interferometer to generate an initial hybrid-superposition of OAM and polarization. Based on the circular birefringence property of light beams propagating in $^{87}$Rb atomic vapor when a longitudinal magnetic field is applied, we produced a magnetic-field-induced rotation of the interference pattern of helical light. We measured the dependence of the rotation angle on magnetic field intensity. Moreover, we determined the background magnetic field intensity by the symmetrical nature of the dispersion-like shape curve. This study offers a method to perform magnetic-field sensing.

This work was supported by the National Fundamental Research Program of China (Grant No. 2011CBA00200) and the National Natural Science Foundation of China (Grant Nos. 11174271, 61275115, and 61435011).